\documentclass[11pt]{article}
\usepackage{moriond,epsfig,graphics}
\newcommand{\pythjet}{\textsc{Py\-thia/Jet\-set}}
\newcommand{\cerenkov}{\v{C}erenkov}

\newcommand{\dn}{\ensuremath{D^{\raise0.3ex\hbox{\scriptsize{\;\!0}}}}}

\newcommand{\eg}{e.g.}

\newcommand{\etal}{{\em et al.}}

\def\issue(#1,#2,#3){{\bf #1} (#3) #2} 

\def\opcit(#1){ {\em op. cit.}, #1}

\def\ARNPS(#1,#2,#3){{\em Ann.\ Rev.\ Nucl.\ Part.\ Sci.} \issue(#1,#2,#3)}
\def\CPC(#1,#2,#3){{\em Comp.\ Phys.\ Comm.} \issue(#1,#2,#3)}
\def\CIP(#1,#2,#3){{\em Comput.\ Phys.} \issue(#1,#2,#3)}
\def\CJP(#1,#2,#3){{\em Chin.\ J.\ Phys.\ (Taipai)} \issue(#1,#2,#3)}
\def\EPJC(#1,#2,#3){{\em Eur.\ Phys.\ J.}  \issue(#1,#2,#3)}
\def\IEEETNS(#1,#2,#3){{\em IEEE Trans.\ Nucl.\ Sci.} \issue(#1,#2,#3)}
\def\MPL(#1,#2,#3){{\em Mod.\ Phys.\ Lett.} \issue(#1,#2,#3)}
\def\NP(#1,#2,#3){{\em Nucl.\ Phys.} \issue(#1,#2,#3)}
\def\NIM(#1,#2,#3){{\em  Nucl.\ Inst.\ and Meth.} \issue(#1,#2,#3)}
\def\PL(#1,#2,#3){{\em Phys.\ Lett.} \issue(#1,#2,#3)}
\def\PRD(#1,#2,#3){{\em Phys.\ Rev.} \issue(#1,#2,#3)}
\def\PRL(#1,#2,#3){{\em Phys.\ Rev.\ Lett.} \issue(#1,#2,#3)}
\def\ZPC(#1,#2,#3){{\em Zeit.\ Phys.} C \issue(#1,#2,#3)}

\def\be{\begin{equation}}
\def\ee{\end{equation}}
\def\bea{\begin{eqnarray}}
\def\eea{\end{eqnarray}}

\begin{document}
\begin{table}[t]
\begin{center}
\tabcolsep=10.8mm
\begin{tabular}{ll} 
XXXV Rencontre de Moriond &  UMS/HEP/2000-030 \\ 
Electroweak Interactions and Unified Theories  & FERMILAB-Conf-00/223-E \\
Les Arcs, France (11--18 March 2000)  & \\ 
\end{tabular}
\end{center}
\end{table}
\vskip 16pt
\vspace*{11mm}

\title{SEARCH FOR RARE AND FORBIDDEN CHARM MESON DECAYS AT FERMILAB E791}

\author{D.~J.~Summers,$^1$
E.~M.~Aitala,$^1$
S.~Amato,$^2$
J.~C.~Anjos,$^2$
J.~A.~Appel,$^6$
D.~Ashery,$^{14}$
S.~Banerjee,$^6$
I.~Bediaga,$^2$
G.~Blaylock,$^9$
S.~B.~Bracker,$^{15}$
P.~R.~Burchat,$^{13}$
R.~A.~Burnstein,$^7$
T.~Carter,$^6$
H.~S.~Carvalho,$^2$
N.~K.~Copty,$^{12}$
L.~M.~Cremaldi,$^1$
C.~Darling,$^{18}$
K.~Denisenko,$^6$
S.~Devmal,$^4$
A.~Fernandez,$^{11}$
G.~F.~Fox,$^{12}$
P.~Gagnon,$^3$
C.~Gobel,$^2$
K.~Gounder,$^1$
A.~M.~Halling,$^6$
G.~Herrera,$^5$
G.~Hurvits,$^{14}$
C.~James,$^6$
P.~A.~Kasper,$^7$
S.~Kwan,$^6$
D.~C.~Langs,$^{12}$
J.~Leslie,$^3$
B.~Lundberg,$^6$
J.~Magnin,$^2$
S.~MayTal-Beck,$^{14}$
B.~Meadows,$^4$
J.~R.~T.\ de Mello Neto,$^2$
D.~Mihalcea,$^8$
R.~H.~Milburn,$^{16}$
J.~M.~de~Miranda,$^2$
A.~Napier,$^{16}$
A.~Nguyen,$^8$
A.~B.~d'Oliveira,$^{4,\,11}$
K.~O'Shaughnessy,$^3$
K.~C.~Peng,$^7$
L.~P.~Perera,$^4$
M.~V.~Purohit,$^{12}$
B.~Quinn,$^1$
S.~Radeztsky,$^{17}$
A.~Rafatian,$^1$
N.~W.~Reay,$^8$
J.~J.~Reidy,$^1$
A.~C.~dos Reis,$^2$
H.~A.~Rubin,$^7$
D.~A.~Sanders,$^1$
A.~K.~S.~Santha,$^4$
A.~F.~S.~Santoro,$^2$
A.~J.~Schwartz,$^4$
M.~Sheaff,$^{5,\,17}$
R.~A.~Sidwell,$^8$
A.~J.~Slaughter,$^{18}$
M.~D.~Sokoloff,$^4$
J.~Solano,$^2$
N.~R.~Stanton,$^8$
R.~J.~Stefanski,$^6$
K.~Stenson,$^{17}$  
S.~Takach,$^{18}$
K.~Thorne,$^6$
A.~K.~Tripathi,$^8$
S.~Watanabe,$^{17}$
R.~Weiss-Babai,$^{14}$
J.~Wiener,$^{10}$
N.~Witchey,$^8$
E.~Wolin,$^{18}$
S.~M.~Yang,$^8$
D.~Yi,$^1$
S.~Yoshida,$^8$
R.~Zaliznyak,$^{13}$ and
C.~Zhang$^8$}

\address{
\begin{table}[h]
\begin{center}
\tabcolsep=4.0pt
\begin{tabular}{ll} 
\em \small $^1$Univ.\ of Mississippi, Oxford, MS 38677, USA &
\em \small $^2$CBPF, Rio de Janeiro, Brazil \\
\em \small $^3$Univ.\ of California, Santa Cruz, CA 95064, USA &
\em \small $^4$Univ.\ of Cincinnati, Cincinnati, OH 45221, USA \\
\em \small $^5$CINVESTAV, 07000 Mexico City, DF Mexico &
\em \small $^6$Fermilab, Batavia, IL 60510, USA \\
\em \small $^7$Illinois Institute of Tech., Chicago, IL 60616, USA &
\em \small $^8$Kansas State Univ., Manhattan, KS 66506, USA \\
\em \small $^9$Univ.\ of Massachusetts, Amherst, MA 01003, USA &
\em \small $^{10}$Princeton University, Princeton, NJ 08544, USA \\
\em \small $^{11}$Universidad Autonoma de Puebla, Mexico &
\em \small $^{12}$Univ.\ of South Carolina, Columbia, SC 29208, USA \\
\em \small $^{13}$Stanford University, Stanford, CA 94305, USA &
\em \small $^{14}$Tel Aviv University, Tel Aviv 69978, Israel \\
\em \small $^{15}$Box 1290, Enderby, BC \ V0E 1V0, Canada &
\em \small $^{16}$Tufts University, Medford, MA 02155, USA \\
\em \small $^{17}$Univ.\ of Wisconsin, Madison, WI 53706, USA &
\em \small $^{18}$Yale University, New Haven, CT 06511, USA \\
\end{tabular}
\end{center}
\end{table}
}
\vspace*{121.3mm}
\leftline{\hspace{66.2mm}
\resizebox{35mm}{!}{\includegraphics{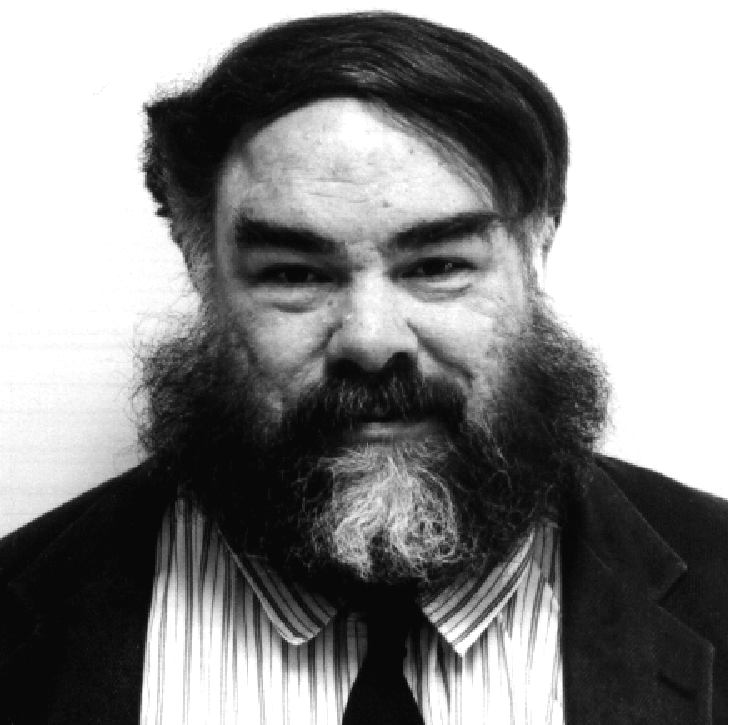}}}
\vspace*{-159.3mm}
\maketitle\abstracts{
We report the results of a {\it blind} search for flavor-changing neutral 
current, 
lepton-flavor violating, and lepton-number violating decays of
$D^+$, $D_{s}^{+}$, and $D^0$ mesons (and their antiparticles) into 
modes containing muons and electrons. Using data from Fermilab charm 
hadroproduction experiment E791, we examine the $\pi \ell \ell$ and 
$K\ell \ell$ decay modes of $D^+$ and $D_{s}^{+}$ and the 
$\ell^+ \ell^-$ decay modes of $D^0$. No evidence for any of these 
decays is found. Therefore, we present branching-fraction upper limits 
at 90$\%$ confidence level for the 24 decay modes examined. Eight of 
these modes have no previously reported limits, and fourteen are 
reported with significant improvements over previously published 
results.}

\newpage
\begin{table}[t]
\begin{center}
\tabcolsep=10.8mm
\begin{tabular}{ll} 
XXXV Rencontre de Moriond &  UMS/HEP/2000-030 \\ 
Interactions \'Electrofaibles et Th\'eories Unifi\'ees. 
& FERMILAB-Conf-00/223-E \\
Les Arcs, France (11--18 mars 2000)  & \\ 
\end{tabular}
\end{center}
\end{table}
\vskip 16pt
\vspace*{11mm}
\title{RECHERCHE DE D\'ESINT\'EGRATIONS DE M\'ESONS CHARM\'ES PAR MODES
RARES ET/OU INTERDITS PAR L'EXP\'ERIENCE E791 \`A FERMILAB}

\author{D.~J.~Summers,$^1$
E.~M.~Aitala,$^1$
S.~Amato,$^2$
J.~C.~Anjos,$^2$
J.~A.~Appel,$^6$
D.~Ashery,$^{14}$
S.~Banerjee,$^6$
I.~Bediaga,$^2$
G.~Blaylock,$^9$
S.~B.~Bracker,$^{15}$
P.~R.~Burchat,$^{13}$
R.~A.~Burnstein,$^7$
T.~Carter,$^6$
H.~S.~Carvalho,$^2$
N.~K.~Copty,$^{12}$
L.~M.~Cremaldi,$^1$
C.~Darling,$^{18}$
K.~Denisenko,$^6$
S.~Devmal,$^4$
A.~Fernandez,$^{11}$
G.~F.~Fox,$^{12}$
P.~Gagnon,$^3$
C.~Gobel,$^2$
K.~Gounder,$^1$
A.~M.~Halling,$^6$
G.~Herrera,$^5$
G.~Hurvits,$^{14}$
C.~James,$^6$
P.~A.~Kasper,$^7$
S.~Kwan,$^6$
D.~C.~Langs,$^{12}$
J.~Leslie,$^3$
B.~Lundberg,$^6$
J.~Magnin,$^2$
S.~MayTal-Beck,$^{14}$
B.~Meadows,$^4$
J.~R.~T.\ de Mello Neto,$^2$
D.~Mihalcea,$^8$
R.~H.~Milburn,$^{16}$
J.~M.~de~Miranda,$^2$
A.~Napier,$^{16}$
A.~Nguyen,$^8$
A.~B.~d'Oliveira,$^{4,\,11}$
K.~O'Shaughnessy,$^3$
K.~C.~Peng,$^7$
L.~P.~Perera,$^4$
M.~V.~Purohit,$^{12}$
B.~Quinn,$^1$
S.~Radeztsky,$^{17}$
A.~Rafatian,$^1$
N.~W.~Reay,$^8$
J.~J.~Reidy,$^1$
A.~C.~dos Reis,$^2$
H.~A.~Rubin,$^7$
D.~A.~Sanders,$^1$
A.~K.~S.~Santha,$^4$
A.~F.~S.~Santoro,$^2$
A.~J.~Schwartz,$^4$
M.~Sheaff,$^{5,\,17}$
R.~A.~Sidwell,$^8$
A.~J.~Slaughter,$^{18}$
M.~D.~Sokoloff,$^4$
J.~Solano,$^2$
N.~R.~Stanton,$^8$
R.~J.~Stefanski,$^6$
K.~Stenson,$^{17}$  
S.~Takach,$^{18}$
K.~Thorne,$^6$
A.~K.~Tripathi,$^8$
S.~Watanabe,$^{17}$
R.~Weiss-Babai,$^{14}$
J.~Wiener,$^{10}$
N.~Witchey,$^8$
E.~Wolin,$^{18}$
S.~M.~Yang,$^8$
D.~Yi,$^1$
S.~Yoshida,$^8$
R.~Zaliznyak,$^{13}$ et
C.~Zhang$^8$}

\address{
\begin{table}[h]
\begin{center}
\tabcolsep=4.0pt
\begin{tabular}{ll} 
\em \small $^1$Univ.\ of Mississippi, Oxford, MS 38677, USA &
\em \small $^2$CBPF, Rio de Janeiro, Brazil \\
\em \small $^3$Univ.\ of California, Santa Cruz, CA 95064, USA &
\em \small $^4$Univ.\ of Cincinnati, Cincinnati, OH 45221, USA \\
\em \small $^5$CINVESTAV, 07000 Mexico City, DF Mexico &
\em \small $^6$Fermilab, Batavia, IL 60510, USA \\
\em \small $^7$Illinois Institute of Tech., Chicago, IL 60616, USA &
\em \small $^8$Kansas State Univ., Manhattan, KS 66506, USA \\
\em \small $^9$Univ.\ of Massachusetts, Amherst, MA 01003, USA &
\em \small $^{10}$Princeton University, Princeton, NJ 08544, USA \\
\em \small $^{11}$Universidad Autonoma de Puebla, Mexico &
\em \small $^{12}$Univ.\ of South Carolina, Columbia, SC 29208, USA \\
\em \small $^{13}$Stanford University, Stanford, CA 94305, USA &
\em \small $^{14}$Tel Aviv University, Tel Aviv 69978, Israel \\
\em \small $^{15}$Box 1290, Enderby, BC \ V0E 1V0, Canada &
\em \small $^{16}$Tufts University, Medford, MA 02155, USA \\
\em \small $^{17}$Univ.\ of Wisconsin, Madison, WI 53706, USA &
\em \small $^{18}$Yale University, New Haven, CT 06511, USA \\
\end{tabular}
\end{center}
\end{table}
}
\vspace*{120.7mm}
\leftline{\hspace{66.2mm}
\resizebox{35mm}{!}{\includegraphics{summers00.eps}}}
\vspace*{-158.7mm}
\maketitle\abstracts{
Les r\'esultats d'une recherche {\em aveugle} portant sur des courants
neutres de changement de saveur ou des violations de la conservation
de la saveur ou du nombre leptonique sont pr\'esent\'ees \`a partir de
l'\'etude de d\'esint\'egrations des m\'esons charm\'es $D^+$, $D_{s}^{+}$,
et $D^0$  ainsi que leur antiparticules via des modes contenant soit des
\'electrons, soit des muons.
Bas\'e sur l'\'echantillon de donn\'ees
amass\'ees par l'exp\'erience d'hadroproduction de charme E791 \`a Fermilab,
nous examinons les modes de d\'esint\'egration de $D^+$ et $D_{s}^{+}$ via
$\pi \ell \ell$ et $K\ell \ell$ ainsi que $D^0 \to \ell^+ \ell^-$.
Aucune \'evidence pour ces types de d\'esint\'egration n'a \'et\'e trouv\'ee.
Nous d\'erivons donc des limites sup\'erieures correspondant
\`a des intervalles de confiance de 90\% pour les 24 modes examin\'es.
Huit de ces limites n'avaient jamais \'et\'e mesur\'ees au
pr\'ealable et quatorze autres repr\'esentent une am\'elioration
consid\'erable sur les limites ant\'erieures.}

\newpage
One way to discover physics beyond the Standard Model is to search for
decays that are forbidden or else are predicted to occur at a negligible
level. If seen, such decays might require new physics such as the introduction
of a new particle to
mediate the decays.  
Many experiments have examined decays of the charge 1/3 strange and beauty 
quarks.\cite{strange}
Here, we look for rare and forbidden decays involving the charge 2/3 charm
quark.  Charge 2/3 quarks may couple differently than 
charge 1/3 quarks.\cite{Pakvasa}

We present the results of a search\,\cite{Sanders} for 24 decay modes of 
charmed $D$ mesons and their antiparticles. 
These decay modes
fall into three categories:
\vspace*{-3pt}
\begin{enumerate}
\item FCNC -- flavor-changing neutral current decays 
($D^0\rightarrow\ell^+\ell^-$ and
$D^+_{(d,s)}\rightarrow h^+\ell^+\ell^-$); 
\item LFV -- lepton-flavor violating decays 
($D^{0}\rightarrow \mu^{\pm }e^{\mp }$,
$D^+_{(d,s)}\rightarrow h^+\mu^{\pm }e^{\mp }$, and 
$D^+_{(d,s)}\rightarrow h^-\mu^+e^+$,
in which the leptons belong to different generations and $h$ is
$\pi$ or $K$);
\item LNV -- lepton-number violating decays
($D^+_{(d,s)}\rightarrow h^-\ell^+\ell^+$,
in which the leptons belong to the same generation
but have the same sign charge).
\end{enumerate}
\vspace*{-3pt}
Decay modes belonging to (1) occur within the Standard Model
via higher-order diagrams, but the branching fractions are
at the $10^{-6}$ to $10^{-8}$ level,\cite{SCHWARTZ93}  
below current sensitivity. However, if 
additional particles such as squarks or charginos exist, 
they could contribute additional amplitudes that would make these modes 
observable. Decays in (2) or (3) do not conserve lepton 
number and thus are forbidden. However, lepton 
number conservation is not required by Lorentz or gauge 
invariance, and a number of theoretical extensions to the Standard 
Model predict lepton-number violation.\cite{Pakvasa} 
The limits we present here for rare and forbidden dilepton 
decays of the $D$ mesons are typically more stringent than those 
obtained from previous searches, \cite{E791FCNC,E687,E653,E771,E789} 
or else are the first
reported.

The data are from Fermilab E791,\cite{e791spect} which 
recorded $2 \times 10^{10}$ events at up to 10 MBytes/s.\cite{DA791}
These events were produced by a 500 GeV/$c$~ $\pi ^{-}$ beam 
in five target foils.
Track and vertex reconstruction were provided by 23 silicon microstrip 
planes\,\cite{SMD} and 45 wire chamber planes, plus two magnets.

Electron identification (ID) was based on 
transverse shower shape plus the match of tracks
to shower positions and energies in our electromagnetic 
calorimeter.\cite{SLIC}
ID efficiency varied from 62$\%$  
below 9 GeV to 45$\%$ above 20 GeV. 
The probability to 
mis-ID a pion as an electron
was about 
0.8$\%$.

Muon ID was obtained from two planes of scintillation 
counters. The first plane (5.5m $\times$ 3.0m) of 14 counters measured the
horizontal $x$ axis while the second plane (3.0m $\times$ 2.2m) of 16 
counters measured the vertical $y$ axis.
The counters had 
15 interaction lengths of shielding.
Candidate muon tracks 
were required to pass cuts that were 
set using
$D^+\rightarrow \overline{K}^{*0} \mu^{+}\nu _{\!_{\mu}}$ decays from 
our data.\cite{Chong} Timing from the $y$ 
counters was used to improve the $x$ position resolution. 
Counter efficiencies were 
measured using muons originating from the primary beam 
dump, and were found to be $(99\pm 1)\%$ for the $y$ counters 
and $(69\pm 3)\%$ for the $x$ counters. The probability for 
misidentifying a pion as a muon decreased with momentum; from 
about 6$\%$ at 8 GeV/$c$ to $(1.3 \pm 0.1)\%$ above 
20 GeV/$c$.

After reconstruction of our 50 Terabyte data set,\cite{FARM791} 
events with evidence of well-separated 
production (primary) and decay (secondary) vertices were selected
to separate charm candidates from background. 
Secondary and primary vertices had to be separated 
by more than $20\,\sigma_{_{\!L}}$ for $D^+$ decays and 
more than $12\,\sigma_{_{\!L}}$ for $D^0$ and $D_{s}^{+}$ decays, 
where $\sigma_{_{\!L}}$ is the calculated longitudinal resolution.
The secondary vertex had to be 
separated from the closest material in the target foils by more than 
$5\,\sigma_{_{\!L}}^{\prime }$,
where $\sigma_{_{\!L}}^{\prime }$ is
the separation uncertainty.
The sum of the vector momenta of 
the tracks from the secondary vertex was required to pass within 
$40~\mu$m of the primary vertex. 
Finally, the net momentum of the charm candidate transverse to 
the line connecting the production and decay vertices had to be 
less than 300, 250, and 200 MeV/$c$ for $D^0$, $D_{s}^{+}$, and $D^+$
candidates, respectively.
These cuts 
and our \cerenkov{}\,\cite{Bartlett}
kaon ID cuts were the same 
for each search mode and for its normalization mode.

We used a {\em blind} analysis technique. Before cuts 
were finalized, all events within a 
mass window $\Delta M_S$ around the mass of the $D^{+}$, 
$D_{s}^{+}$, or $D^{0}$ were {\em masked} so that the presence or 
absence of any potential signal would not bias our choice of 
cuts. All cuts were chosen by studying 
signal events generated by a Monte Carlo simulation program (see 
below) and background events from real data. Events within the signal 
windows were unmasked only after this optimization. Background events 
were chosen from a mass window $\Delta M_B$ above and below the signal 
window $\Delta M_S$. The cuts were chosen to maximize the ratio 
$N_S/\sqrt{N_B}$, where $N_S$ and $N_B$ are the numbers of signal and 
background events, respectively. We used asymmetric windows for the 
decay modes containing electrons to allow for the bremsstrahlung 
low-energy tail. The signal windows are: 
\begin{table}[h!]
\begin{center}
\renewcommand{\arraystretch}{1.05}
\vspace*{-6pt}
\begin{tabular}{ll} 
$1.84<M(D^{+})<1.90 ~{\rm for}~ D^{+}\rightarrow h\mu \mu$ \hfil &
$1.78<M(D^{+})<1.90 ~{\rm GeV}/c^{\,2}~{\rm for}~ D^{+}\rightarrow hee ~{\rm
and}~  h\mu e$ \\
$1.95<M(D_{s}^{+})<1.99 ~{\rm for}~ D_{s}^{+}\rightarrow
h\mu \mu$ \hfil &
$1.91<M(D_{s}^{+})<1.99 ~{\rm GeV}/c^{\,2}~{\rm for}~ 
D_{s}^{+}\rightarrow hee~{\rm and}~ h\mu e$ \\
$1.83<M(D^{0})<1.90 ~{\rm for}~ 
D^{0}\rightarrow \mu \mu$ \hfil &
$1.76<M(D^{0})<1.90 ~{\rm GeV}/c^{\,2}~{\rm for}~ 
D^{0}\rightarrow ee~{\rm and}~ \mu e$ \\
\end{tabular}
\vspace*{-5pt}
\end{center}
\end{table}

\begin{figure}[ht]
\vskip -1.7 cm
\centerline{\epsfxsize 5.0 truein \epsfbox{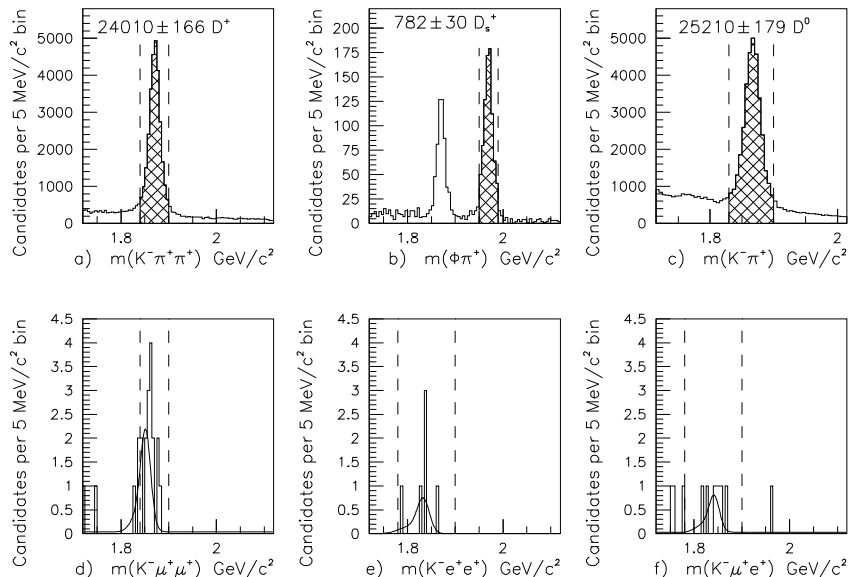}}
\vskip -.2 cm
\vspace*{-11pt}
\caption[]{
\small 
Top row: typical normalization charm signals. 
The signal region is 
shaded. 
Bottom row: invariant mass plots of $D^{+}$ candidate decays to 
$K^-\mu ^+\mu ^+$, $K^-e^+e^+$, and $K^-\mu ^+e^+$, showing 
reflections mostly from misidentified $D^+\rightarrow 
K^-\pi^+\pi^+$ decays. These modes are used to set mis-ID rate
rather than upper limits 
The solid curves are normalized 
Monte Carlo fits. The dashed lines show the signal window.
}
\label{Norm}
\end{figure}

We normalize the sensitivity of our search to topologically similar 
Cabibbo-favored decays. For the $D^{+}$ decays we use 
$D^+\rightarrow K^-\pi^+\pi^+$; for $D_{s}^{+}$ we use 
$D_{s}^{+}\rightarrow \phi \pi^+$; and for $D^{0}$ we 
use $D^0\rightarrow K^-\pi^+$. The mass widths of our normalization modes 
were 10.5 MeV/$c^{\,2}$ for $D^{+}$, 9.5 MeV/$c^{\,2}$ for $D_{s}^{+}$, 
and 12 MeV/$c^{\,2}$ for $D^{0}$. The events within the 
$\sim 5\,\sigma $ window are shown in Figs. \ref{Norm}a--c. 
The upper limit for each branching fraction is $B_{X} =  
({N_{X}}/{N_{\mathrm{Norm}}}) \cdot
({\varepsilon _{\mathrm{Norm}}}/{\varepsilon _{X}})
\cdot B_{\mathrm{Norm}}$, 
where $N_{X}$ is the 90$\%$ CL upper limit on the number of decays 
for the rare or forbidden decay mode $X$, and $\varepsilon_{X}$ is that 
mode's detection efficiency. $N_{\mathrm{Norm}}$ is the fitted number 
of normalization mode decays; $\varepsilon_{\mathrm{Norm}}$ 
is the normalization mode detection efficiency; and 
$B_{\mathrm{Norm}}$ is the normalization mode branching fraction.\cite{PDG} 

The ratio of detection efficiencies is given by 
${\varepsilon _{\mathrm{Norm}}}/{\varepsilon _{X}} =
{N_{\mathrm{Norm}}^{\mathrm{MC}}}/{N_{X}^{\mathrm{MC}}}$,
where $N_{\mathrm{Norm}}^{\mathrm{MC}}$ and $N_{X}^{\mathrm{MC}}$ are 
the fractions of Monte Carlo events that are reconstructed and pass 
final cuts, for the normalization and decay modes, 
respectively. We use \pythjet~\cite{MC} as the physics 
generator and model the effects of resolution, geometry, magnetic 
fields, multiple scattering, interactions in the detector material, 
detector efficiencies, and the analysis cuts. The 
efficiencies for the normalization modes varied from about 
$0.5\%$ to $2\%$ and for the 
search modes varied from about $0.1\%$ to $2\%$. 

Monte Carlo studies show that the experiment's acceptances are nearly 
uniform across the Dalitz plots, except that the dilepton 
ID efficiencies typically drop to near zero at the 
dilepton mass threshold. 
The efficiency typically reaches its full value at masses 
only a few hundred MeV/$c^{\,2}$ above the dilepton mass threshold. We 
use a constant weak-decay matrix element when calculating the overall 
detection efficiencies. Two exceptions to the use of the Monte Carlo 
simulations in determining relative efficiencies are made: those for 
\cerenkov{} ID when the number of kaons in the signal and 
normalization modes are different, and those for the muon 
ID. These efficiencies are determined from data.

\begin{figure}[t!]
\vskip -2.5 cm
\centerline{\epsfxsize 5.0 truein \epsfbox{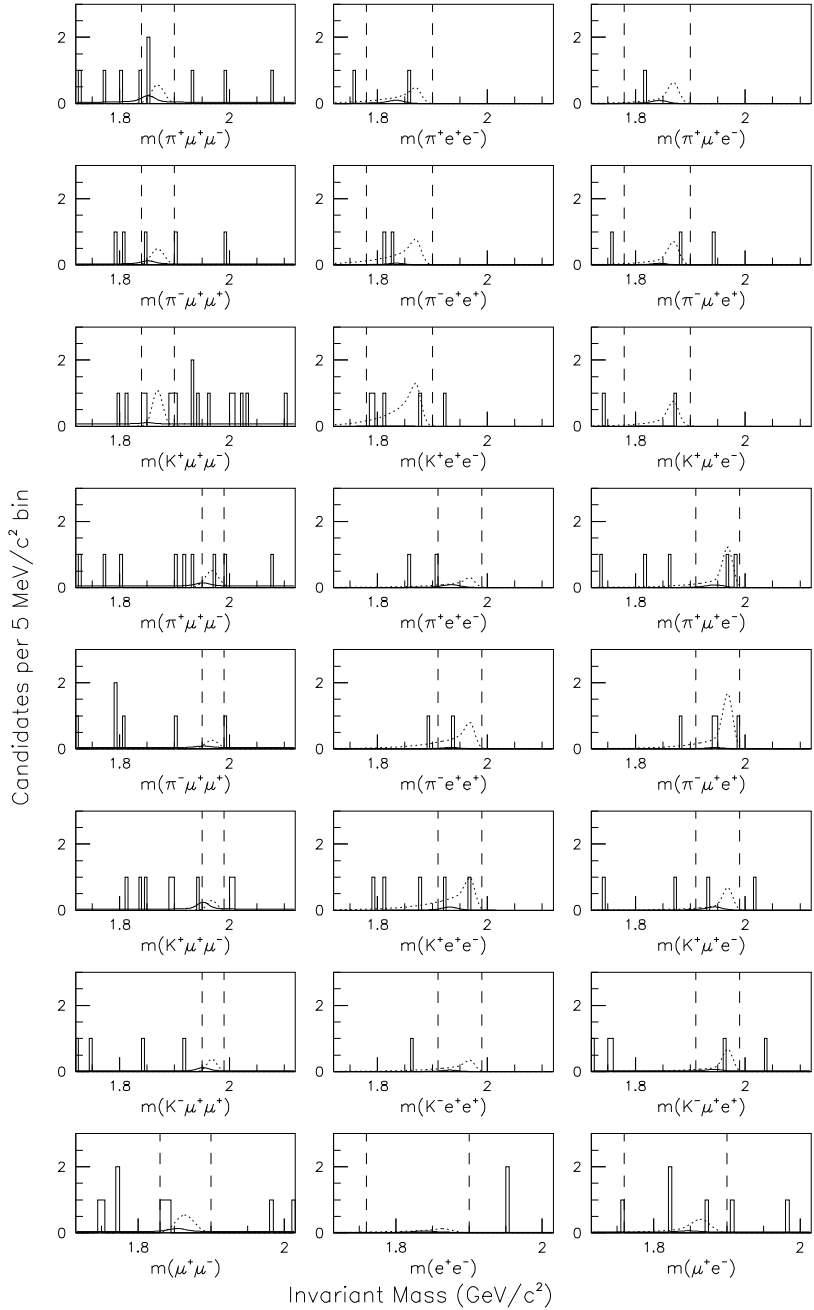}}
\vskip -.2 cm
\vspace*{-11pt}
\caption[]{
\small Final event samples for the $D^+$ (rows 1--3), 
$D_{s}^{+}$ (rows 4--7), and $D^0$ (row 8) decays. The solid curves 
represent estimated background; the dotted curves represent signal 
shape for a number of events equal to the 90$\%$ CL upper limit. 
The dashed vertical lines are $\Delta M_S$ boundaries.}
\label{Data}
\vspace*{-12pt}
\end{figure}
\begin{table}[b!]
\vskip -8mm
\caption[]{
E791 90$\%$ confidence level (CL) branching fractions (BF) compared 
to previous experiments. 
The background and candidate events 
correspond to the signal region only.}
\label{Results}
\vskip 5pt
\tabcolsep=4.0pt
\begin{center}
\begin{tabular}{lccccclll} \hline 
&(Est.&BG)&Cand.&Syst.&90$\%$ CL&E791&Previous & Previous  \\
Mode&$N_{\mathrm{Cmb}}$&$N_{\mathrm{MisID}}$&Obs.&Err.&Num.
&$BF$ Limit&$BF$ Limit & Experiment \\
\hline
\vspace*{-11pt} &     &   &     &     &         &    &         &     \\
 $D^{+}\rightarrow \pi ^{+}\mu ^{+}\mu ^{-}$&1.20&1.47&2&10$\%$&3.35
 &$1.5\times 10^{-5}$&$1.8\times 10^{-5}$ & E791 \cite{E791FCNC} \\
 $D^{+}\rightarrow \pi ^{+}e^{+}e^{-}$&0.00&0.90&1&12$\%$&3.53
 &$5.2\times 10^{-5}$&$6.6\times 10^{-5}$ & E791 \cite{E791FCNC} \\
 $D^{+}\rightarrow \pi ^{+}\mu ^{\pm }e^{\mp }$&0.00&0.78&1&11$\%$&3.64
 &$3.4\times 10^{-5}$&$1.2\times 10^{-4}$ & E687 \cite{E687} \\
 $D^{+}\rightarrow \pi ^{-}\mu ^{+}\mu ^{+}$&0.80&0.73&1&9$\%$&2.92
 &$1.7\times 10^{-5}$&$8.7\times 10^{-5}$ & E687 \cite{E687} \\
 $D^{+}\rightarrow \pi ^{-}e^{+}e^{+}$&0.00&0.45&2&12$\%$&5.60
 &$9.6\times 10^{-5}$&$1.1\times 10^{-4}$ & E687 \cite{E687} \\
 $D^{+}\rightarrow \pi ^{-}\mu ^{+}e^{+}$&0.00&0.39&1&11$\%$&4.05
 &$5.0\times 10^{-5}$&$1.1\times 10^{-4}$ & E687 \cite{E687} \\
 $D^{+}\rightarrow K^{+}\mu ^{+}\mu ^{-}$&2.20&0.20&3&8$\%$&5.07
 &$4.4\times 10^{-5}$&$9.7\times 10^{-5}$ & E687 \cite{E687} \\
 $D^{+}\rightarrow K^{+}e^{+}e^{-}$&0.00&0.09&4&11$\%$&8.72
 &$2.0\times 10^{-4}$&$2.0\times 10^{-4}$ & E687 \cite{E687} \\
 $D^{+}\rightarrow K^{+}\mu ^{\pm }e^{\mp }$&0.00&0.08&1&9$\%$&4.34
 &$6.8\times 10^{-5}$&$1.3\times 10^{-4}$ & E687 \cite{E687} \\
\hline
\vspace*{-11pt} &     &   &     &     &         &    &         &     \\
 $D_{s}^{+}\rightarrow K^{+}\mu ^{+}\mu ^{-}$&0.67&1.33&0&27$\%$&1.32
 &$1.4\times 10^{-4}$&$5.9\times 10^{-4}$ & E653 \cite{E653} \\
 $D_{s}^{+}\rightarrow K^{+}e^{+}e^{-}$&0.00&0.85&2&29$\%$&5.77
 &$1.6\times 10^{-3}$& & \\
 $D_{s}^{+}\rightarrow K^{+}\mu ^{\pm }e^{\mp }$&0.40&0.70&1&27$\%$&3.57
 &$6.3\times 10^{-4}$& & \\
 $D_{s}^{+}\rightarrow K^{-}\mu ^{+}\mu ^{+}$&0.40&0.64&0&26$\%$&1.68
 &$1.8\times 10^{-4}$&$5.9\times 10^{-4}$  & E653 \cite{E653} \\
 $D_{s}^{+}\rightarrow K^{-}e^{+}e^{+}$&0.00&0.39&0&28$\%$&2.22
 &$6.3\times 10^{-4}$& & \\
 $D_{s}^{+}\rightarrow K^{-}\mu ^{+}e^{+}$&0.80&0.35&1&27$\%$&3.53
 &$6.8\times 10^{-4}$& & \\
 $D_{s}^{+}\rightarrow \pi ^{+}\mu ^{+}\mu ^{-}$&0.93&0.72&1&27$\%$&3.02
 &$1.4\times 10^{-4}$&$4.3\times 10^{-4}$  & E653 \cite{E653} \\
 $D_{s}^{+}\rightarrow \pi ^{+}e^{+}e^{-}$&0.00&0.83&0&29$\%$&1.85
 &$2.7\times 10^{-4}$& & \\
 $D_{s}^{+}\rightarrow \pi ^{+}\mu ^{\pm }e^{\mp }$&0.00&0.72&2&30$\%$
 &6.01&$6.1\times 10^{-4}$& & \\
 $D_{s}^{+}\rightarrow \pi ^{-}\mu ^{+}\mu ^{+}$&0.80&0.36&0&27$\%$&1.60
 &$8.2\times 10^{-5}$&$4.3\times 10^{-4}$  & E653 \cite{E653} \\
 $D_{s}^{+}\rightarrow \pi ^{-}e^{+}e^{+}$&0.00&0.42&1&29$\%$&4.44
 &$6.9\times 10^{-4}$& &\\
 $D_{s}^{+}\rightarrow \pi ^{-}\mu ^{+}e^{+}$&0.00&0.36&3&28$\%$&8.21
 &$7.3\times 10^{-4}$& & \\
\vspace*{-12pt} &     &   &     &     &         &    &         &     \\
\hline
\vspace*{-11pt} &     &   &     &     &         &    &         &     \\
 $D^{0}\rightarrow \mu ^{+}\mu ^{-}$&1.83&0.63&2&6$\%$&3.51
 &$5.2\times 10^{-6}$&$4.1\times 10^{-6}$  & BEATRICE \cite{E771} \\
 $D^{0}\rightarrow e^{+}e^{-}$&1.75&0.29&0&9$\%$&1.26
 &$6.2\times 10^{-6}$&$8.2\times 10^{-6}$ & E789 \cite{E789} \\
 $D^{0}\rightarrow \mu ^{\pm }e^{\mp }$&2.63&0.25&2&7$\%$&3.09
 &$8.1\times 10^{-6}$&$1.7\times 10^{-5}$  & E789 \cite{E789} \\
\hline
\end{tabular}
\end{center}
\end{table}

The 90$\%$ CL upper limits $N_{X}$ are calculated using the method of 
Feldman and Cousins \cite{Cousins} to account for background, and then 
corrected for systematic errors by the method of Cousins and 
Highland.\cite{COUSINSHI} In these methods, the numbers of signal events are
determined by simple counting, not by a fit. All results are listed in 
Table \ref{Results} and shown in Fig. \ref{Data}. The 
kinematic criteria and removal of reflections (see below) are different 
for the $D^{+}$, $D_{s}^{+}$, and $D^0$. Thus, the $D^+$ and 
$D_{s}^{+}$ rows in Fig. \ref{Data} with the same decay particles are 
different, and the seventh row of Fig. \ref{Data} is different from the 
bottom row of Fig. \ref{Norm}.

The upper limits are determined by both the number of candidate events 
and the expected number of background events within the signal region.  
Background that is not removed by cuts 
include decays in which hadrons (from 
real, fully-hadronic decay vertices) are misidentified as leptons. 
In the case where kaons are misidentified as leptons, candidates 
have effective masses which lie outside the signal windows. Most of 
these originate from Cabibbo-favored modes 
$D^+\rightarrow K^-\pi^+\pi^+$, $D_{s}^{+}\rightarrow K^-K^+\pi^+$, 
and $D^0\rightarrow K^-\pi^+$. These 
Cabibbo-favored reflections were explicitly removed prior to 
cut optimization. There remain two sources of background 
in our data: hadronic decays with pions misidentified as leptons 
($N_{\mathrm{MisID}}$) and ``combinatoric'' background 
($N_{\mathrm{Cmb}}$) arising primarily from false vertices and 
partially reconstructed charm decays. After cuts 
were applied and the signal windows opened, the number of events 
within the window is $N_{\mathrm{Obs}} = N_{\mathrm{Sig}}+ 
N_{\mathrm{MisID}} + N_{\mathrm{Cmb}}$. 

The background $N_{\mathrm{MisID}}$ arises mainly 
from singly-Cabibbo-suppressed (SCS) modes. These misidentified 
leptons can come from hadronic shower punchthrough, 
decays-in-flight, and random overlaps of 
tracks. We do 
not attempt to establish a limit for $D^+\rightarrow K^-\ell^+\ell^+$ 
modes, as they have relatively large feedthrough signals from copious 
Cabibbo-favored $K^-\pi^+\pi^+$ decays. Instead, we use the observed 
signals in $K^-\ell^+\ell^+$ channels to measure three dilepton 
mis-ID rates under the assumption that the observed signals 
(shown in Figs.~\ref{Norm}d--f) arise entirely from lepton 
mis-ID. The curve shapes are from Monte Carlo. 
The following mis-ID rates were obtained: 
$r_{\mu\mu}= (7.3 \pm 2.0)\times 10^{-4}$, 
$r_{\mu e}= (2.9 \pm 1.3 )\times 10^{-4}$, and 
$r_{e e}= (3.4 \pm 1.4)\times 10^{-4}$. 
Using these rates we estimate the numbers of misidentified candidates, 
$N_{\mathrm{MisID}}^{h\ell\ell}$ (for $D^{+}$ and $D_{s}^{+}$) and 
$N_{\mathrm{MisID}}^{\ell\ell}$ (for $D^{0}$), in the signal windows 
as follows:
$N_{\mathrm{MisID}}^{h\ell\ell} = r_{\ell\ell} 
\cdot N_{\mathrm{SCS}}^{h\pi\pi}$
and
$N_{\mathrm{MisID}}^{\ell\ell} = r_{\ell\ell} 
\cdot N_{\mathrm{SCS}}^{\pi\pi}$,
where $N_{\mathrm{SCS}}^{h\pi\pi}$ and $N_{\mathrm{SCS}}^{\pi\pi}$ 
are the numbers of SCS hadronic decay candidates within the signal 
windows. For modes in which two possible pion combinations can 
contribute, \eg, $D^+\rightarrow h^{+}\mu ^{\pm}\mu ^{\mp}$, we double 
the rate. 

To estimate the combinatoric background $N_{\mathrm{Cmb}}$ within a 
signal window $\Delta M_S$, we count events having masses within an 
adjacent background mass window $\Delta M_B$, and scale this number 
($N_{\Delta M_B}$) by the relative sizes of these windows:
$N_{\mathrm{Cmb}} = ({\Delta M_S}/{\Delta M_B}) \cdot N_{\Delta M_B}$. 
To be conservative in calculating our 90$\%$ confidence level upper 
limits, we take combinatoric backgrounds to be zero when no 
events are located above the mass windows. In Table \ref{Results} we 
present the numbers of combinatoric background, mis-ID 
background, and observed events for all 24 modes.

Systematic errors in this analysis include: 
statistical errors from the fit to the normalization sample 
$N_{\mathrm{Norm}}$; statistical errors on the numbers of Monte Carlo 
events for both $N_{\mathrm{Norm}}^{\mathrm{MC}}$ and 
$N_{X}^{\mathrm{MC}}$; uncertainties in the calculation of 
mis-ID background; and uncertainties in the relative 
efficiency for each mode, including lepton and kaon tagging. 
These tagging efficiency uncertainties include: 1) the 
muon counter efficiencies from both Monte Carlo simulation and 
hardware performance; 2) kaon \cerenkov{} ID 
efficiency due to differences in kinematics and modeling between 
data and Monte Carlo simulated events; and 3) the fraction of 
signal events (based on simulations) that would remain outside the 
signal window due to bremsstrahlung tails. 
The larger systematic errors for the $D_{s}^{+}$ modes, 
compared to the $D^{+}$ and $D^{0}$ modes, are due to the uncertainty 
in the branching fraction for the $D_{s}^{+}$ normalization mode.
The sums, taken in quadrature, of these systematic errors are listed 
in Table \ref{Results}.

In summary, we use a {\em blind} analysis of data from Fermilab 
E791 to obtain upper limits on the dilepton branching 
fractions for flavor-changing neutral current, lepton-number violating, 
and lepton-family violating decays of $D^+$, $D_{s}^{+}$, and $D^0$ 
mesons. No evidence for any of these decays is found. 
The 90$\%$ 
confidence level branching fraction limits shown in Table \ref{Results}
represent significant improvements over 
previously published results. 
In the future we hope to report results for 4-prong decays of the
$D^0$ charm meson to a pair of leptons and either a neutral vector 
meson \cite{Singer}
or a $\pi\pi$, $\pi K$, or $KK$ pair.

This research was supported by the U.S.\ DOE and NSF,
the Brazilian Conselho Nacional de Desenvolvimento Cient\'\i fico e
Tecnol\'ogico, CONACyT (Mexico), the Israeli Academy of Sciences and
Humanities, and the U.S.-Israel Binational
Science Foundation.

\end{document}